\documentclass[journal]{IEEEtran}

\usepackage{amsmath}
\usepackage{algorithm}
\usepackage{algorithmic}
\usepackage{booktabs}
\usepackage{multirow}
\usepackage{cite}
\usepackage[utf8]{inputenc} 
\usepackage[T1]{fontenc}    
\usepackage{hyperref}       
\usepackage{url}            

\usepackage{amsfonts}       
\usepackage{nicefrac}       
\usepackage{microtype}      
\usepackage{xcolor}         
\usepackage{graphicx}
\usepackage{float}
\usepackage[numbers,sort&compress]{natbib}
\usepackage{caption}

\usepackage{xcolor}

\definecolor{IssueGeneral}{RGB}{0,102,204}
\definecolor{IssueLength}{RGB}{204,102,0}
\definecolor{IssueStability}{RGB}{180,40,40}
\definecolor{IssueUncertainty}{RGB}{120,70,160}

\newfloat{figtab}{htb}{fgtb}
\makeatletter
  \newcommand\figcaption{\def\@captype{figure}\caption}
  \newcommand\tabcaption{\def\@captype{table}\caption}
\makeatother

\def\BibTeX{{\rm B\kern-.05em{\sc i\kern-.025em b}\kern-.08em
    T\kern-.1667em\lower.7ex\hbox{E}\kern-.125emX}}

\title{Learning Dynamic Neural Evidence Representations for Time-Adaptive Brain–Computer Interfaces
}

\author{
\IEEEauthorblockN{Beining Cao\textsuperscript{1}, Ziyi Zhao\textsuperscript{1}, Xiaowei Jiang\textsuperscript{1}, Daniel Leong\textsuperscript{1}, Yingtao Ren \textsuperscript{1}, Thomas Do\textsuperscript{1}\textsuperscript{*}, Yu-Cheng Fred Chang\textsuperscript{1}, Chin-Teng Lin\textsuperscript{1}}\\

\IEEEauthorblockA{
\textsuperscript{1} Australian AI Institute, School of Computer Science,\\ Faculty of Engineering and Information Technology, University of Technology Sydney}

\thanks{\textsuperscript{*}Corresponding author: Chin-Teng Lin. Email: Chin-Teng.Lin@uts.edu.au}

}

\begin{document}
\maketitle

\begin{abstract}

Brain–computer interfaces (BCIs) decode neural activity into commands, yet most existing systems rely on fixed-window decoding that may result in redundant observation or unreliable predictions due to insufficient evidence. Adaptive temporal decision-making (ATDM) addresses this accuracy–time trade-off by progressively accumulating EEG evidence and deciding when to stop. However, existing EEG encoders are mainly designed for fixed-window decoding and may not provide reliable state representations under variable observation lengths. In addition, current ATDM-oriented encoders are typically tailored to specific EEG paradigms, limiting their applicability across different BCI tasks. To address these limitations, we propose ProtoTrigger, a two-stage prototype learning-based EEG state encoder for ATDM. ProtoTrigger uses prototype matching to extract stable local EEG embeddings and prototype-based attention to aggregate decision-relevant temporal evidence during progressive observation. Offline evaluations across three EEG paradigms demonstrated state-of-the-art accuracy-time trade-offs and strong generalizability across different EEG paradigms. An online human-in-the-loop augmented reality-based BCI experiment further demonstrated its real-time feasibility.
 These results suggest that ProtoTrigger provides a general EEG state encoding framework for efficient ATDM-based BCI systems.
\end{abstract}

\begin{IEEEkeywords}
Brain-computer interface, EEG, Adaptive temporal decision-making, Prototype learning, Human-in-the-loop
\end{IEEEkeywords}

\section{Introduction}

\begin{figure}[h!]
  \centering
  \includegraphics[width=1\linewidth]{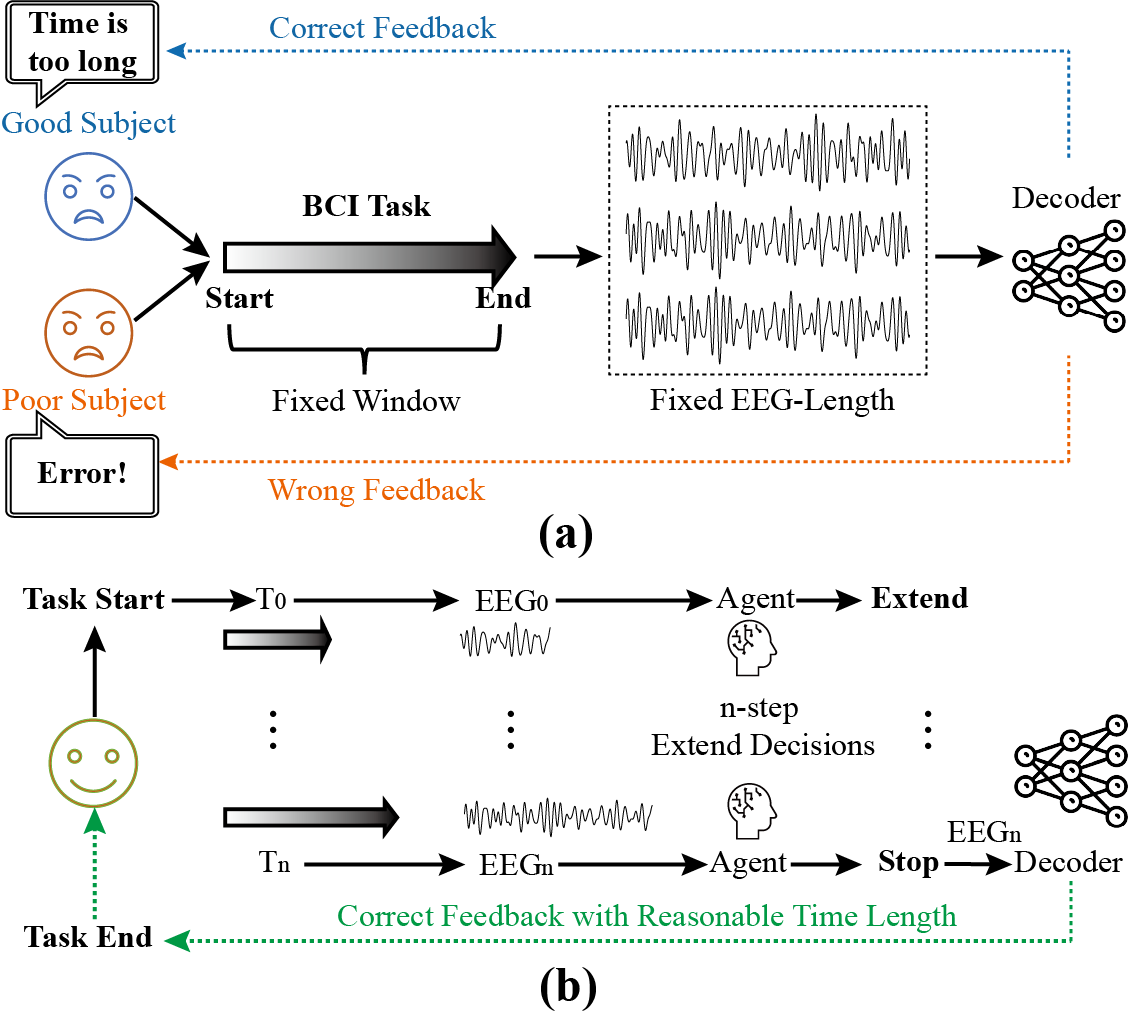}

\caption{Illustration of (a) FW-based BCI and (b) ATDM-based BCI. In ATDM-based BCI, the decision policy adaptively determines the EEG observation duration based on progressively accumulated evidence.}
\label{fig:Figure1}
\end{figure}

\IEEEPARstart{E}{lectroencephalography} (EEG)-based brain--computer interfaces (BCIs) constitute an intuitive neural interaction modality for human--machine interaction and embodied intelligence \cite{lu2026motor,jiang2025ifuzzytl}. Representative EEG paradigms, such as steady-state visual evoked potentials (SSVEPs) \cite{wang2026enhancing} and motor imagery (MI) \cite{zhao2026dg}, have been widely adopted in BCI applications. Currently, most BCI systems adopt a fixed-window (FW) strategy, in which decoding is performed after a predefined EEG observation window \cite{lawhern2018eegnet, song2022eeg,nguyen2025edge}. As illustrated in Fig.~\ref{fig:Figure1}(a), FW decoding overlooks inter-subject differences in the required EEG observation duration, potentially resulting in redundant observation or insufficient EEG evidence and compromising the BCI user experience \cite{zhou2024dynamic}.

As illustrated in Fig.~\ref{fig:Figure1}(b), adaptive temporal decision-making (ATDM) dynamically determines the stopping time based on the accumulated EEG evidence, enabling timely and reliable feedback in human-in-the-loop (HITL) BCI systems \cite{yang2020dynamic,zhou2024dynamic}. Specifically, ATDM can be formulated as a reinforcement learning (RL)-based Markov decision process (MDP), where EEG-derived states guide reward-aware temporal decisions \cite{zhuang2026neuro}. 
Accordingly, temporally comparable representations of variable-length EEG observations are desirable for reliable policy evaluation and value estimation \cite{yau2021evidence}. Moreover, state representations should preserve decision-relevant information and reflect prediction uncertainty during progressive evidence accumulation \cite{dabney2021value,ghosh2020representations}.

However, most existing EEG encoders were developed for FW decoding rather than ATDM state construction. Although deep neural encoders, such as CNNs and transformers, can learn discriminative EEG representations \cite{lawhern2018eegnet,he2023classification,song2022eeg}, they do not explicitly promote temporal comparability across progressively accumulated observations, which may affect reliable policy learning in ATDM.
Consequently, encoded states may exhibit cross-length drift as the discriminative quality of EEG evidence varies over time within a trial, potentially reducing the reliability of extend-or-stop decisions. Existing ATDM-oriented EEG encoders are predominantly based on canonical correlation analysis (CCA) for SSVEP BCIs \cite{zhou2024dynamic,yang2020dynamic}. These methods represent EEG states using correlations with predefined frequency-specific templates, providing a fixed comparison basis across progressively extended observation windows. However, these predefined references are paradigm-specific and cannot be generalized across different EEG tasks.

Prototype learning offers a potential solution by replacing manually designed references with learnable prototypes \cite{snell2017prototypical}. This strategy has been explored in EEG transfer learning and domain adaptation, where shared prototypes capture cross-subject neural patterns and mitigate distribution shifts \cite{han2025spatial,jiang2026ifuzz}. In prototype learning, variable-length EEG observations are represented by their similarities to a shared set of learnable prototypes. This mapping establishes a common reference space for aligning EEG evidence across different observation lengths. Prototype-based representations may therefore facilitate the construction of temporally comparable states for ATDM.

To enable general EEG state encoding for ATDM, we propose ProtoTrigger, a prototype learning-based EEG state encoder. The encoder consists of two complementary prototype-based stages. Local temporal-spatial features are extracted by matching EEG patches to learnable prototypes, while prototype-based attention assigns adaptive importance to evidence from different temporal segments during progressive observation. This design supports state encoding from progressively accumulated EEG observations while adaptively aggregating decision-relevant temporal evidence. The proposed method is evaluated under RL-based ATDM settings on datasets from three distinct EEG paradigms, achieving superior accuracy-time trade-offs. The practical feasibility of the proposed ProtoTrigger-based ATDM BCI system is further validated through an HITL online experiment.

The contributions of this study are summarized as follows:
\begin{enumerate}
\item \textbf{Prototype learning-Based EEG State Encoding:} A two-stage prototype learning-based EEG state encoder is proposed for learning reliable state representations from variable-length observations across different EEG paradigms in ATDM-based BCIs.
\item \textbf{Cross-Paradigm Offline Validation:} ProtoTrigger is evaluated across three EEG paradigms, demonstrating state-of-the-art accuracy--time trade-offs and strong generalizability across different paradigms.
\item \textbf{Online Feasibility Validation:} ProtoTrigger is deployed in a real-time HITL BCI system, demonstrating improved accuracy--time trade-off over FW strategies with negligible computational latency.
\end{enumerate}

\section{Related Work}

\subsection{Existing EEG Encoding Methods}

EEG encoding aims to derive discriminative temporal-spatial representations from neural signals \cite{lawhern2018eegnet}. Existing EEG encoders can be broadly divided into knowledge-driven methods and data-driven deep neural methods. Knowledge-driven encoders incorporate paradigm-specific signal-processing priors. In SSVEP BCI, where decoding relies primarily on frequency-related temporal responses, CCA and its variants \cite{chen2015filter, nakanishi2017enhancing} encode EEG signals through correlations with predefined frequency-specific reference templates. In MI BCI, where discriminative information is reflected in spatial activation responses, common spatial pattern (CSP) and its variants encode EEG signals by projecting them onto class-discriminative spatial filters estimated from training data \cite{ramoser2000optimal,ang2008filter}. These temporal templates and spatial filters serve as task-specific reference structures that support stable EEG representations across observation lengths. However, their reliance on strong paradigm-specific priors limits generalizability across different EEG paradigms.

\begin{figure*}[h!]
  \centering
  \includegraphics[width=1\linewidth]{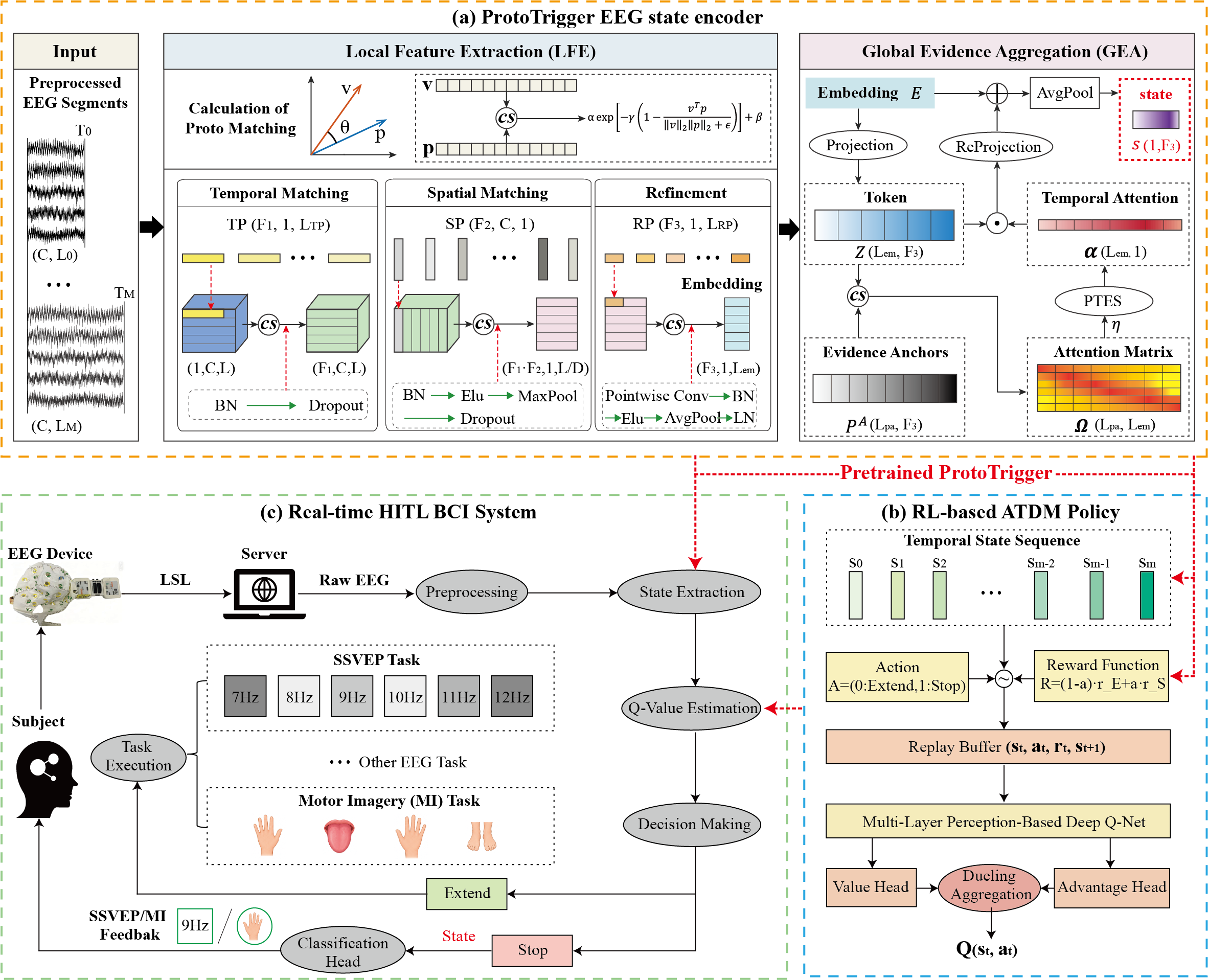}

\caption{Overall architecture of the ProtoTrigger-based ATDM BCI system.
(a) Structure of the ProtoTrigger EEG state encoder. 
(b) RL-based decision-making policy. 
(c) Real-time HITL BCI framework.
In this figure, black arrows indicate the data flow, while red dashed arrows indicate inter-module calls.}
\label{fig:Figure2}
\end{figure*}

By contrast, deep neural encoders learn EEG representations from data. CNN-based models such as EEGNet \cite{lawhern2018eegnet} and TCNet \cite{ingolfsson2020eeg} extract local temporal-spatial patterns through convolutional filters. Transformer-based methods, including EEG-Transformer \cite{he2023classification}, SSVEPFormer \cite{chen2023transformer}, and Conformer \cite{song2022eeg}, further capture global dependencies through self-attention. These data-driven encoders are applicable across diverse EEG paradigms. However, they are generally optimized for FW decoding rather than the temporally comparable state construction required by ATDM. As observation length changes, variations in temporal context and evidence quality may affect the latent state representations, while temporal comparability is not explicitly ensured. Overall, knowledge-driven encoders offer task-specific reference structures but limited generalizability, whereas conventional deep encoders provide broader applicability without explicit alignment to ATDM state requirements. These limitations highlight the need for an EEG state encoder specifically designed for variable-length state construction in ATDM-based BCIs.

\subsection{Prototype Learning in EEG Studies}

Prototype learning bridges knowledge-driven and data-driven EEG encoding by representing observations through their similarities to shared learnable prototypes. In EEG research, it has been applied to transfer learning, domain adaptation, and robust decoding. Lian et al. proposed P2CSL using subspace class prototypes to improve cross-subject EEG classification \cite{lian2025p2csl}. Han et al. developed a dual-prototype framework for small-sample MI representation learning \cite{han2025spatial}. In SSVEP BCI research, fuzzy-prototype-based transfer learning has also been explored for cross-subject adaptation \cite{jiang2025ifuzzytl,jiang2026ifuzz}. Prototype mechanisms have further been introduced for unseen-target emotion recognition \cite{zhou2022pr,li2026pl}, robust MI decoding \cite{cai2024multi}, EEG eye-state classification \cite{nilashi2023electroencephalography}, and few-shot sleep staging \cite{li2023few}.

Existing EEG studies indicate that prototype learning can improve robustness and transferability by capturing shared neural structures across heterogeneous data distributions. By anchoring progressively accumulated EEG observations to shared learnable prototypes, prototype learning may support stable and temporally comparable state construction while preserving evolving evidence for ATDM. However, current prototype learning-based EEG studies have primarily focused on FW-based tasks, leaving variable-length state encoding for ATDM-based BCI largely unexplored.

\section{Method}

Fig.~\ref{fig:Figure2} illustrates the proposed ProtoTrigger-based ATDM BCI framework. Variable-length EEG observations are first preprocessed and encoded into fixed-dimensional states through local feature extraction (LFE) and global evidence aggregation (GEA), after which an RL-based policy performs adaptive stop-or-extend decision-making.

\subsection{Preprocessing}

A single EEG trial is denoted as $\mathbf{X}\in\mathbb{R}^{C\times L}$, where $C$ and $L$ represent the numbers of channels and time points, respectively. To emulate progressive evidence accumulation for model training, each trial is segmented into variable-length observations from $T_0$ to $T_M$ with an increment of $\Delta T$. Butterworth band-pass filtering and channel-wise z-score normalization are applied before encoder and policy training.

\subsection{ProtoTrigger EEG State Encoder}
As shown in Fig.~\ref{fig:Figure2}(a), the proposed ProtoTrigger encoder consists of two modules. First, the LFE module maps each EEG segment to a sequence of local embeddings through prototype matching. Next, the GEA module applies temporal attention across local embeddings and aggregates them into an evidence-enhanced state for decision-making.

\subsubsection{Local Feature Extraction}

As illustrated in Fig.~\ref{fig:Figure2}(a), the LFE module maps the variable-length observation into a sequence of local embeddings through temporal matching, spatial matching, and refinement.
This three-stage architecture follows the compact temporal-spatial feature extraction principle \cite{lawhern2018eegnet}. The prototype-matching-based feature extraction in LFE is motivated by reference-based EEG encoding, which has demonstrated feasibility for ATDM through similarity-based EEG representation \cite{yang2020dynamic}. Accordingly, learnable temporal and spatial prototypes serve as data-adaptive references for EEG state encoding. Similarity-based prototype matching maps local observations to a common space, supporting temporally comparable state construction across observation lengths.

For prototype matching, given an EEG observation
$\mathbf{X}\in\mathbb{R}^{C\times L}$,
a local receptive-field patch indexed by $u$ is vectorized as
$\mathbf{v}_{u}\in\mathbb{R}^{d}$.
A set of $K$ learnable prototypes is denoted by
$\mathcal{P}=\{\mathbf{p}_{k}\}_{k=1}^{K}$.
The prototype-matching operator
$\Psi_{\mathcal{P}}(\cdot,\cdot)$ is defined as
\begin{equation}
\label{eq:prototype_matching}
\Psi_{\mathcal{P}}
\left(
\mathbf{v}_{u},
\mathbf{p}_{k}
\right)
=
\alpha_k
\exp
\left[
-\gamma_k
\left(
1-
\frac{
\mathbf{v}_{u}^{\top}\mathbf{p}_{k}
}{
\|\mathbf{v}_{u}\|_2
\|\mathbf{p}_{k}\|_2
+
\epsilon
}
\right)
\right]
+
\beta_k
\end{equation}
where $\gamma_k$ controls response sharpness, $\alpha_k$ and $\beta_k$ are learnable scaling and bias parameters, respectively. $\epsilon$ is a small constant for numerical stability. A larger response indicates greater similarity between the local EEG patch and the prototype.

First, temporal prototype matching is performed independently on each EEG channel using $F_1$ prototypes of length $L_{\mathrm{TP}}$, denoted by
$\mathcal{TP}\in\mathbb{R}^{F_1\times 1 \times L_{\mathrm{TP}}}$.
For channel $c$ at temporal position index $u$, the local patch is defined as
\begin{equation}
\mathbf{v}^{\mathrm{T}}_{c,u}
=
\operatorname{vec}
\left(
\mathbf{X}_{c,\,u:u+L_{\mathrm{TP}}-1}
\right)
\in
\mathbb{R}^{L_{\mathrm{TP}}}
\end{equation}
where $u$ indexes the sliding temporal position. The normalized response of $\mathbf{v}^{\mathrm{T}}_{c,u}$ to the $k$-th temporal prototype
$\mathbf{p}^{\mathrm{T}}_k\in\mathcal{TP}$ is
\begin{equation}
H^{\mathrm{T}}_{k,c,u}
=
\mathrm{BN}
\left(
\Psi_{\mathcal{TP}}
\left(
\mathbf{v}^{\mathrm{T}}_{c,u},
\mathbf{p}^{\mathrm{T}}_k
\right)
\right)
\qquad
\mathbf{H}^{\mathrm{T}}
\in
\mathbb{R}^{F_1\times C\times L}
\end{equation}

Same padding is applied during temporal prototype matching with a stride of one, preserving the temporal dimension $L$.
This stage extracts local temporal patterns, including frequency-related EEG activity. 

Next, spatial prototype matching is performed to extract spatial features.
For each temporal position, a channel-wise patch
$\mathbf{v}^{\mathrm{S}}\in\mathbb{R}^{1\times C}$
is extracted along the channel dimension from $\mathbf{H}^{\mathrm{T}}$
and matched with spatial prototypes
$\mathcal{SP}\in\mathbb{R}^{F_2\times C\times1}$.
The spatial feature
$\mathbf{H}^{\mathrm{S}}\in\mathbb{R}^{F_1F_2\times1\times L_{\mathrm{S}}}$
is obtained as
\begin{equation}
\mathbf{H}^{\mathrm{S}}
=
\mathrm{MaxPool}_{D}
\left(
\mathrm{ELU}
\left(
\mathrm{BN}
\left(
\Psi_{\mathcal{SP}}
\left(
\mathbf{H}^{\mathrm{T}}
\right)
\right)
\right)
\right)
\end{equation}
where $D$ denotes the temporal pooling factor. This stage encodes inter-channel response relationships.

Finally, refinement prototype matching is applied along the temporal dimension.
For each response map of $\mathbf{H}^{\mathrm{S}}$, a local temporal patch
$\mathbf{v}^{\mathrm{R}}\in\mathbb{R}^{1\times L_{\mathrm{RP}}}$
is extracted over a sliding window and matched with refinement prototypes
$\mathcal{RP}\in\mathbb{R}^{F_3\times1\times L_{\mathrm{RP}}}$.
The resulting local embedding sequence
$\mathbf{E}\in\mathbb{R}^{F_3\times L_{\mathrm{em}}}$
is obtained as
\begin{equation}
\mathbf{E}
=
\mathrm{AvgPool}
\left(
\mathrm{ELU}
\left(
\mathrm{BN}
\left(
\mathrm{PConv}
\left(
\Psi_{\mathcal{RP}}
\left(
\mathbf{H}^{\mathrm{S}}
\right)
\right)
\right)
\right)
\right)
\end{equation}
where $\mathrm{PConv}$ denotes a pointwise convolution used to project and fuse channel-wise features.

\subsubsection{Global Evidence Aggregation}

The GEA module adaptively weights and aggregates the local embeddings generated by LFE into a fixed-dimensional evidence-aware state for ATDM. The aggregation enhances embeddings associated with evidence-rich temporal positions while assigning lower weights to less informative positions. Unlike self-attention, which derives embedding weights from pairwise relations within the current observation window, the proposed prototype-based attention evaluates each local embedding against a shared set of learnable evidence prototypes that serve as anchors for temporal evidence assessment. These evidence anchors provide a shared reference for identifying informative evidence as the observation window is progressively extended.

Given the local embedding sequence
$\mathbf{E}$,
a shared linear projection maps the embedding into the anchor-matching space:
\begin{equation}
\mathbf{Z}
=
\phi_{\mathrm{proj}}
\left(
\mathbf{E}
\right)
\qquad
\mathbf{Z}
\in
\mathbb{R}^{L_{\mathrm{em}}\times F_3}
\end{equation}
where $\phi_{\mathrm{proj}}(\cdot)$ denotes a learnable linear projection.
A set of $L_{\mathrm{pa}}$ learnable evidence anchors is represented by
\begin{equation}
\mathbf{P}^{\mathrm{A}}
=
\left[
\mathbf{p}^{\mathrm{A}}_1,
\mathbf{p}^{\mathrm{A}}_2,
\ldots,
\mathbf{p}^{\mathrm{A}}_{L_{\mathrm{pa}}}
\right]^{\top}
\in
\mathbb{R}^{L_{\mathrm{pa}}\times F_3}
\end{equation}
where $\mathbf{p}^{\mathrm{A}}_q\in\mathbb{R}^{1 \times{F_3}}$ denotes the $q$-th evidence anchor.
The matching response between the $l$-th projected embedding $\mathbf{z}_{l}$ and the $q$-th evidence anchor $\mathbf{p}^{\mathrm{A}}_q$ is also computed using the prototype-matching operator defined in Eq.~\eqref{eq:prototype_matching} and normalized across evidence anchors, yielding $\boldsymbol{\Omega}=[\omega_{q,l}]\in\mathbb{R}^{L_{\mathrm{pa}}\times L_{\mathrm{em}}}$.

A shared MLP-based prototype temporal evidence scorer (PTES) estimates the temporal importance of each local embedding from its anchor-response vector. Specifically, the shared PTES maps each $\boldsymbol{\omega}_{l}$ to a scalar evidence score, and the resulting scores across all temporal positions are normalized to obtain the final temporal attention weights:
\begin{equation}
\boldsymbol{\alpha}
=
\operatorname{PTES}
\left(
\boldsymbol{\Omega}^{\top}
\right)
=
\left[
\alpha_1,
\alpha_2,
\ldots,
\alpha_{L_{\mathrm{em}}}
\right]^{\top}
\in
\mathbb{R}^{L_{\mathrm{em}}\times1}
\end{equation}
The attention-weighted tokens are re-projected and fused with the original local embedding sequence:
\begin{equation}
\mathbf{S}
=
\mathbf{E}
+
\phi_{\mathrm{re}}
\left(
\boldsymbol{\alpha}
\odot
\mathbf{Z}
\right)
\qquad
\mathbf{S}
\in
\mathbb{R}^{L_{\mathrm{em}}\times F_3}
\end{equation}
Accordingly, local embeddings associated with evidence-rich temporal positions receive larger attention weights, whereas less informative embeddings are downweighted.

Finally, the enhanced embedding sequence 
$\mathbf{S}\in\mathbb{R}^{L_{\mathrm{em}}\times F_3}$ is averaged over the temporal dimension $L_{\mathrm{em}}$, yielding a fixed-dimensional state 
$\mathbf{s}\in\mathbb{R}^{F_3}$ for ATDM.
GEA employs prototype-based attention to emphasize decision-relevant EEG evidence and construct a unified state for decision-making.

\subsubsection{Cross-Length Supervised Pretraining}

The proposed encoder is pretrained with label supervision over progressively accumulated EEG observations at all decision steps. Given the encoded state $\mathbf{s}_{b,t}$ of the $b$-th trial at decision step $t$, the encoder and an MLP-based prediction head are jointly optimized using
\begin{equation}
\mathcal{L}_{\mathrm{pre}}
=
\frac{1}{B(M+1)}
\sum_{b=1}^{B}
\sum_{t=0}^{M}
\operatorname{CE}
\left(
g_{\mathrm{pred}}(\mathbf{s}_{b,t}),
y_b
\right)
\end{equation}
where $B$ denotes the batch size, $M$ is the maximum decision step, $g_{\mathrm{pred}}(\cdot)$ denotes the prediction head, and $y_b$ is the ground-truth label of the $b$-th trial.

\subsection{Reinforcement Learning Policy for ATDM}

As illustrated in Fig.~\ref{fig:Figure2}(b), ATDM can be formulated as a Markov decision process in which progressively extended EEG observations are encoded as a state sequence \cite{zhou2024dynamic,yang2020dynamic}
\begin{equation}
\mathcal{S}=\{\mathbf{s}_0,\mathbf{s}_1,\ldots,\mathbf{s}_M\}
\end{equation}
where $\mathbf{s}_t$ denotes the EEG state at step $t$. At each step, an action $a_t \in \mathcal{A}=\{0,1\}$ is selected, where $a_t=0$ corresponds to extending the observation, while $a_t=1$ corresponds to outputting the current prediction.
The reward is defined to trade off accuracy and decision time (DT):
\begin{equation}
r_t=(1-a_t)r_E+a_t r_S
\end{equation}
where $r_E<0$ penalizes additional observation time, and
\begin{equation}
r_S=
\begin{cases}
r_{\mathrm{correct}}, & \hat{y}_t = y\\
r_{\mathrm{wrong}}, & \hat{y}_t \neq y
\end{cases}
\end{equation}
The action value is estimated using a dueling deep Q-network (DQN) as:
\begin{equation}
Q(\mathbf{s}_t,a_t)
=
V(\mathbf{s}_t)
+
A(\mathbf{s}_t,a_t)
-
\frac{1}{|\mathcal{A}|}
\sum_{a' \in \mathcal{A}} A(\mathbf{s}_t,a')
\end{equation}
During inference, the optimal action is selected as
\begin{equation}
a_t^{*}=\arg\max_{a \in \mathcal{A}} Q(\mathbf{s}_t,a)
\end{equation}
The current prediction is output when $a_t^{*}=1$ or when the maximum observation length ($T_M$) is reached. Otherwise, the observation is further extended ($a_t^{*}=0$).

\section{Experiment}
\subsection{Experiment Settings}
\subsubsection{Datasets}
To evaluate the proposed ProtoTrigger encoder across different EEG paradigms, experiments were conducted on three datasets, including SSVEP~\cite{nakanishi2015comparison}, MI~\cite{tangermann2012review}, and a self-collected rotation-frequency hybrid VEP (RFH-VEP) dataset. EEG encoding fundamentally relies on temporal-spatial signal representation \cite{lawhern2018eegnet}. SSVEP, MI, and RFH-VEP are characterized by frequency-related temporal dynamics, sensorimotor spatial patterns, and joint temporal-spatial responses, respectively, thereby enabling evaluation of the robustness of learned state representations across multiple EEG feature domains. Furthermore, these paradigms have demonstrated feasibility for real-time decoding, supporting their suitability for validating ATDM-based BCI systems. Detailed dataset descriptions are provided in Supplementary Section~I.

\subsubsection{Baseline Models}

Several conventional EEG encoders were employed as data-driven baselines, including EEGNet, TCNet, Transformer, Conformer, and iFuzzyTL~\cite{lawhern2018eegnet,ingolfsson2020eeg,he2023classification,song2022eeg,jiang2025ifuzzytl}. Although these models were originally developed for FW EEG classification, they can be reformulated as state encoders for the ATDM task. At each decision step, variable-length feature maps were generated from the accumulated EEG by the corresponding encoder and then average-pooled to form a fixed-dimensional state representation $\mathbf{s}_t$.
All data-driven baselines were pretrained with the same loss function as ProtoTrigger.
For comparison with conventional knowledge-driven EEG encoders, an SSVEP-specific FBCCA encoder and an MI-specific CSP encoder were employed for the SSVEP and MI datasets, respectively. A fixed-window ProtoTrigger (FWPT) baseline was also evaluated at successive observation lengths from $T_0$ to $T_M$ with an interval of $\Delta T$. The highest ITR across these observation lengths was reported. All compared encoders shared the same ATDM framework. FWPT used the same pretrained ProtoTrigger encoder but produced predictions at fixed observation lengths.
Detailed baseline settings are provided in Supplementary Section~II.

\subsection{Implementation Details}

\subsubsection{Model Configuration}

EEG signals were first band-pass filtered between 2 and 70~Hz using a Butterworth filter and normalized by channel-wise z-score normalization. The initial observation window was set to 0.5~s for SSVEP and RFH-VEP and 1.0~s for MI, given the delayed onset of motor-intention-related neural activity \cite{pfurtscheller2001motor,lu2026motor}. For all datasets, the observation window was extended in 0.25~s increments up to 4.0~s.

For the ProtoTrigger LFE module, the number and length of the temporal prototypes were set to $F_1=16$ and $L_{\mathrm{TP}}=128$, respectively. The spatial prototype length was set to match the number of selected channels, which was 8, 17, and 22 for SSVEP, MI, and RFH-VEP, respectively. The number of spatial prototypes was fixed at two for all three datasets. The refinement prototype length and final state dimension were set to $L_{\mathrm{RP}}=31$ and $F_3=32$, respectively. The temporal pooling factor $D$ in LFE was set to 4. In the GEA module, $L_{\mathrm{pa}}=24$ learnable evidence anchors were used. The dropout rate was set to 0.5.
The RL agent was parameterized by a three-layer MLP with hidden dimensions of 256, 128, and 64. The discount factor was set to 0.99 and the target network was updated every 200 optimization steps. The rewards for extension, correct stopping, and incorrect stopping were set to $r_E=-0.03$, $r_{\mathrm{correct}}=0.6$, and $r_{\mathrm{wrong}}=-0.4$, respectively. The same DQN configuration was adopted for all three tasks. For all baseline encoders, the same RL configuration as ProtoTrigger was adopted.

\subsubsection{Data Partition}
A five-fold cross-validation protocol was adopted, with three folds used for encoder training, one for encoder validation and subsequent DQN training, and one for testing. The training split was used for encoder pretraining. The validation split was first used for encoder checkpoint selection and was subsequently repurposed for DQN training, while the test split remained completely unseen during both stages. Sliding-window augmentation \cite{song2022eeg} was applied only during encoder and DQN training, whereas all inference trials remained unaugmented. The same trial partitions and augmentation protocol were used for all compared methods and ablation studies.

\subsubsection{Optimization and Hardware Settings}
All experiments were implemented in PyTorch and conducted on an NVIDIA A40 GPU. All encoders were trained with AdamW for 500 epochs using a batch size of 64 and a learning rate of $1\times10^{-4}$. The validation-selected checkpoint was then frozen for RL training. The DQN agent was trained with AdamW for 300 epochs using a batch size of 64 and a learning rate of $2\times10^{-4}$. All optimization settings were held constant across experiments.


\begin{table*}[t]
\centering
\caption{Overall comparison of ITR (bits/min), decoding accuracy (\%), and DT (seconds) under the ATDM setting across the three datasets. ``--'' indicates that the method is not applicable to the corresponding dataset. Superscript stars denote paired $t$-test results relative to ProtoTrigger under the same dataset and metric (\textsuperscript{*}$p<0.05$, \textsuperscript{**}$p<0.01$, and \textsuperscript{***}$p<0.001$).}

{\fontsize{8pt}{9.2pt}\selectfont
\renewcommand{\arraystretch}{1.18}
\setlength{\tabcolsep}{0pt}

\begin{tabular*}{\textwidth}{
@{\extracolsep{\fill}}
l
l l l
@{\hspace{18pt}}
l l l
@{\hspace{18pt}}
l l l
@{}
}
\toprule
\multirow{2}{*}{Method}
& \multicolumn{3}{c}{SSVEP}
& \multicolumn{3}{c}{MI}
& \multicolumn{3}{c}{RFH-VEP} \\
\cmidrule(lr){2-4}
\cmidrule(lr){5-7}
\cmidrule(lr){8-10}
& ITR & ACC & DT
& ITR & ACC & DT
& ITR & ACC & DT \\
\midrule

Conformer
& 128.34\textsuperscript{***} & 79.72\textsuperscript{*}   & 1.19\textsuperscript{*}
& 22.72\textsuperscript{*}    & 64.42\textsuperscript{**}  & 1.57
& 108.39\textsuperscript{**}  & 67.46\textsuperscript{**}  & 1.07 \\

EEGNet
& 110.21\textsuperscript{***} & 75.00\textsuperscript{**}  & 1.24\textsuperscript{*}
& 16.49\textsuperscript{*}    & 60.02\textsuperscript{***} & 1.71
& 119.30\textsuperscript{***} & 70.67\textsuperscript{*}   & 1.05 \\

TCNet
& 60.27\textsuperscript{***}  & 60.55\textsuperscript{***} & 1.36\textsuperscript{*}
& 12.90\textsuperscript{**}   & 52.08\textsuperscript{***} & 1.67
& 94.46\textsuperscript{***}  & 64.98\textsuperscript{***} & 1.10\textsuperscript{*} \\

Transformer
& 152.55                      & 84.89                      & 1.12\textsuperscript{*}
& 19.58\textsuperscript{**}   & 62.30\textsuperscript{**}  & 1.66
& 110.04\textsuperscript{**}  & 69.93\textsuperscript{*}   & 1.06 \\

iFuzzyTL
& 136.37\textsuperscript{**}  & 78.47\textsuperscript{*}   & 1.14\textsuperscript{*}
& 16.52\textsuperscript{**}   & 53.16\textsuperscript{*}   & 1.49
& 106.37\textsuperscript{***} & 72.76                      & 1.16\textsuperscript{*} \\

FBCCA 
& 121.22\textsuperscript{***} & 85.67                      & 1.45\textsuperscript{**}
& --                          & --                         & --
& --                          & --                         & -- \\

CSP 
& --                          & --                         & --
& 22.13\textsuperscript{*}    & 65.60\textsuperscript{*}   & 1.68
& --                          & --                         & -- \\

\textbf{ProtoTrigger}
& \textbf{163.06}             & \textbf{82.28}             & \textbf{0.97}
& \textbf{25.32}              & \textbf{69.13}             & \textbf{1.60}
& \textbf{134.24}             & \textbf{73.49}             & \textbf{0.99} \\

FWPT
& 133.94\textsuperscript{**}  & 81.67\textsuperscript{*}   & 1.11
& 20.60\textsuperscript{**}   & 60.00\textsuperscript{*}   & 1.41
& 104.42\textsuperscript{***} & 70.98\textsuperscript{*}   & 1.10 \\

\bottomrule
\end{tabular*}
\par}
\label{tab:Table1}
\end{table*}

\subsection{Evaluation Metrics}

Information transfer rate (ITR) was adopted as the primary evaluation metric. As a standard BCI metric, ITR is used to jointly assess decoding accuracy and DT, thereby enabling evaluation of the accuracy--time trade-off in ATDM
\cite{zhou2024dynamic, yau2021evidence}. ITR is defined as
\begin{equation}
\mathrm{ITR}
=
\left[
\log_2 N
+
P\log_2 P
+
(1-P)\log_2
\left(
\frac{1-P}{N-1}
\right)
\right]
\frac{60}{T},
\end{equation}
where $N$ denotes the number of target classes, $P$ denotes the decoding accuracy, and $T$ denotes the mean DT in seconds. Specifically, $T$ denotes the mean EEG observation duration at which the stopping decision was made, averaged across all trials. A higher ITR indicates a better trade-off between decoding accuracy and DT.

\section{Results}
The ITR, decoding accuracy, and DT achieved by all compared methods under the ATDM setting across the three datasets are summarized in Table~\ref{tab:Table1}. Detailed individual-subject results are provided in Tables~I--III of the Supplementary. ProtoTrigger achieved the highest mean ITR on all three datasets, reaching 163.06, 25.32, and 134.24 bits/min, with gains of 10.51, 2.60, and 14.94 bits/min over the best baselines. In addition, ProtoTrigger yielded the shortest DT on SSVEP (0.97 s), the highest accuracy on MI (69.13\%), and both the highest accuracy (73.49\%) and the shortest DT (0.99 s) on RFH-VEP. Paradigm-specific methods, including FBCCA and CSP, achieved high accuracy but yielded constrained ITRs due to their longer DT. ProtoTrigger also consistently outperformed FWPT on all three datasets, indicating that the ATDM strategy provides a better accuracy--time trade-off than the FW strategy.

The observed performance differences may be associated with the distinct characteristics of the three paradigms. In SSVEP, stimulus-frequency responses provide rapid discriminative temporal evidence, enabling ProtoTrigger to achieve accurate decisions at shorter observation durations. In MI, where spatial patterns are less stable, the discriminability of the state becomes more critical. ProtoTrigger therefore improved ITR primarily through higher accuracy without a substantial increase in DT.
RFH-VEP involves both temporal evidence accumulation and spatial discrimination, thereby imposing joint requirements on timely evidence utilization and discriminative state construction.
Consequently, ProtoTrigger achieved the best ITR with the highest accuracy and shortest DT on RFH-VEP. These results indicate that ProtoTrigger adapts to different EEG paradigms by enabling early stopping when temporal evidence is sufficient while preserving state discriminability under less stable neural representations.

\section{Discussion}

\subsection{Ablation Study on Prototype-Based Modules}

\begin{table}[t]
\centering
\caption{Ablation-study ITR (bits/min) results for different module variants on three datasets. Superscript stars denote significant paired $t$-test differences from P-P ($p<0.05$).}

{\fontsize{8pt}{9.2pt}\selectfont
\setlength{\tabcolsep}{4pt}
\renewcommand{\arraystretch}{1}

\begin{tabular}{lcccccc}
\toprule
Dataset & C-P & C-T & P-T & P-N & C-N & \textbf{P-P} \\
\midrule
SSVEP
& 153.31
& 147.05\textsuperscript{*}
& 149.59\textsuperscript{*}
& 130.92\textsuperscript{*}
& 104.04\textsuperscript{*}
& \textbf{163.06} \\

MI
& 21.31\textsuperscript{*}
& 21.77\textsuperscript{*}
& 23.62
& 18.67\textsuperscript{*}
& 16.42\textsuperscript{*}
& \textbf{25.32} \\

RFH-VEP
& 117.69\textsuperscript{*}
& 118.67\textsuperscript{*}
& 122.20\textsuperscript{*}
& 106.80\textsuperscript{*}
& 102.58\textsuperscript{*}
& \textbf{134.24} \\
\bottomrule
\label{tab:Table2}
\end{tabular}
}
\end{table}

To evaluate the contribution of the prototype-based designs in ProtoTrigger, an ablation study was conducted on the LFE and GEA modules. In LFE, the prototype-based extractor was compared with a CNN-based extractor following the compact temporal-spatial convolution design of EEGNet \cite{lawhern2018eegnet}. In GEA, the proposed prototype-based attention was compared with Transformer self-attention \cite{he2023classification} and a no-attention setting. Six variants were evaluated, including CNN-prototype (C-P), CNN-Transformer (C-T), CNN-None (C-N), the proposed prototype-prototype (P-P), prototype-Transformer (P-T), and prototype-None (P-N). All variants were evaluated using the same settings as the main experiments.

As shown in Table~\ref{tab:Table2}, the proposed P-P configuration achieved the highest mean ITR across all three datasets. A general ITR reduction was observed when either prototype-based module was replaced. Moreover, removing GEA resulted in significant performance degradation across all datasets ($p<0.05$). On SSVEP, the non-significant difference between C-P and P-P suggests that strong frequency-specific features can be effectively captured by either local extractor, whereas the significant superiority of P-P over P-T and P-N indicates that prototype-based attention facilitates earlier stopping by effectively weighting temporally distributed evidence. Conversely, on MI, the non-significant difference between P-T and P-P may be attributed to weaker sensorimotor patterns, under which discriminative local feature extraction is more essential than temporal evidence aggregation. The proposed prototype-based LFE and GEA appear to provide complementary contributions to ATDM, where LFE is primarily associated with accuracy, whereas GEA is primarily associated with DT optimization.

\subsection{Analysis of State Representation Quality}

\begin{figure}[h!]
  \centering
  \includegraphics[width=0.9\linewidth]{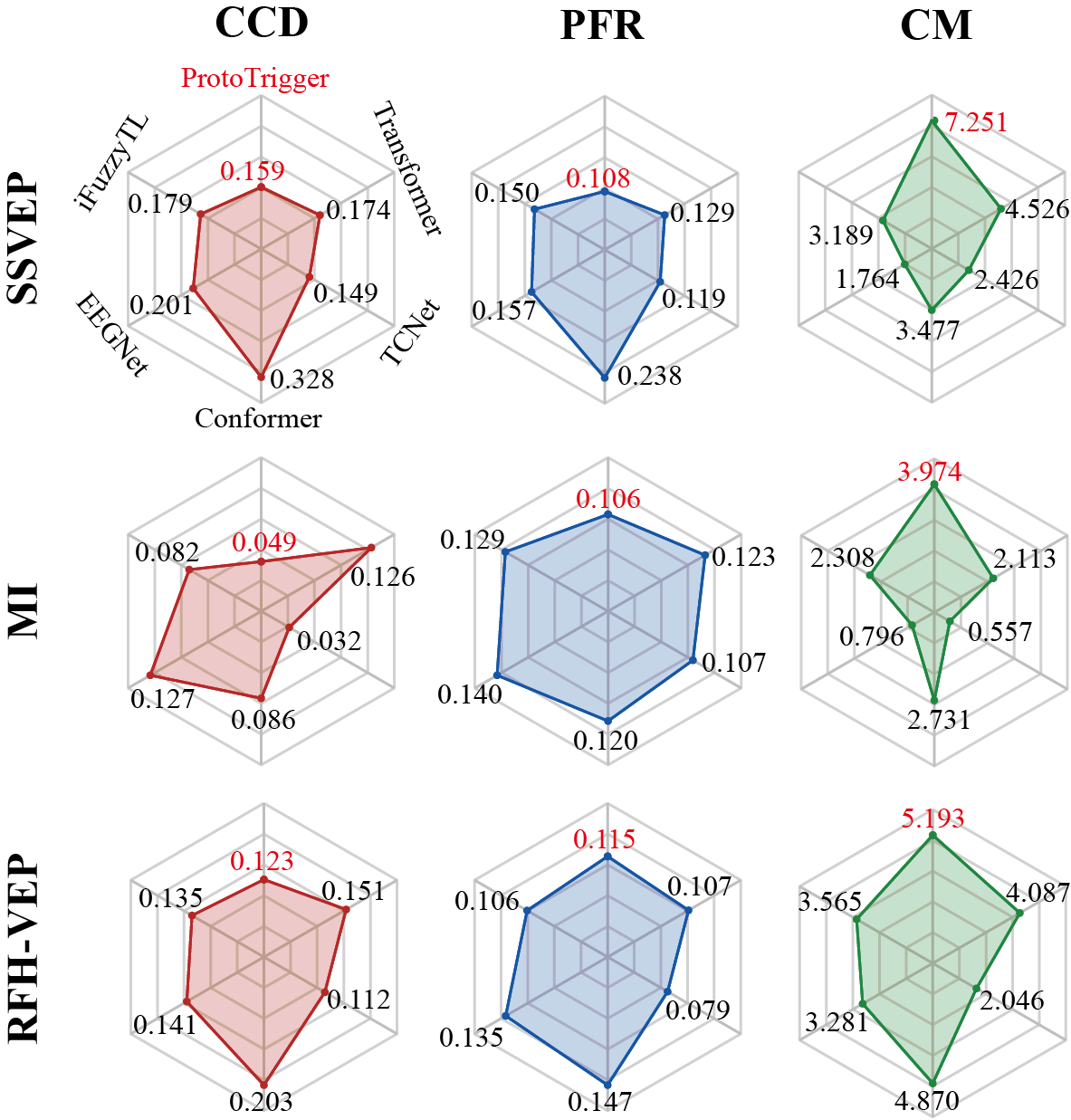}

\caption{
State-representation metrics of different EEG encoders across three datasets. Each polygon vertex denotes a specific method, following the same order shown in subplot (a).}
\label{fig:Figure5}
\end{figure}

\begin{figure*}[h!]
  \centering
  \includegraphics[width=0.8\linewidth]{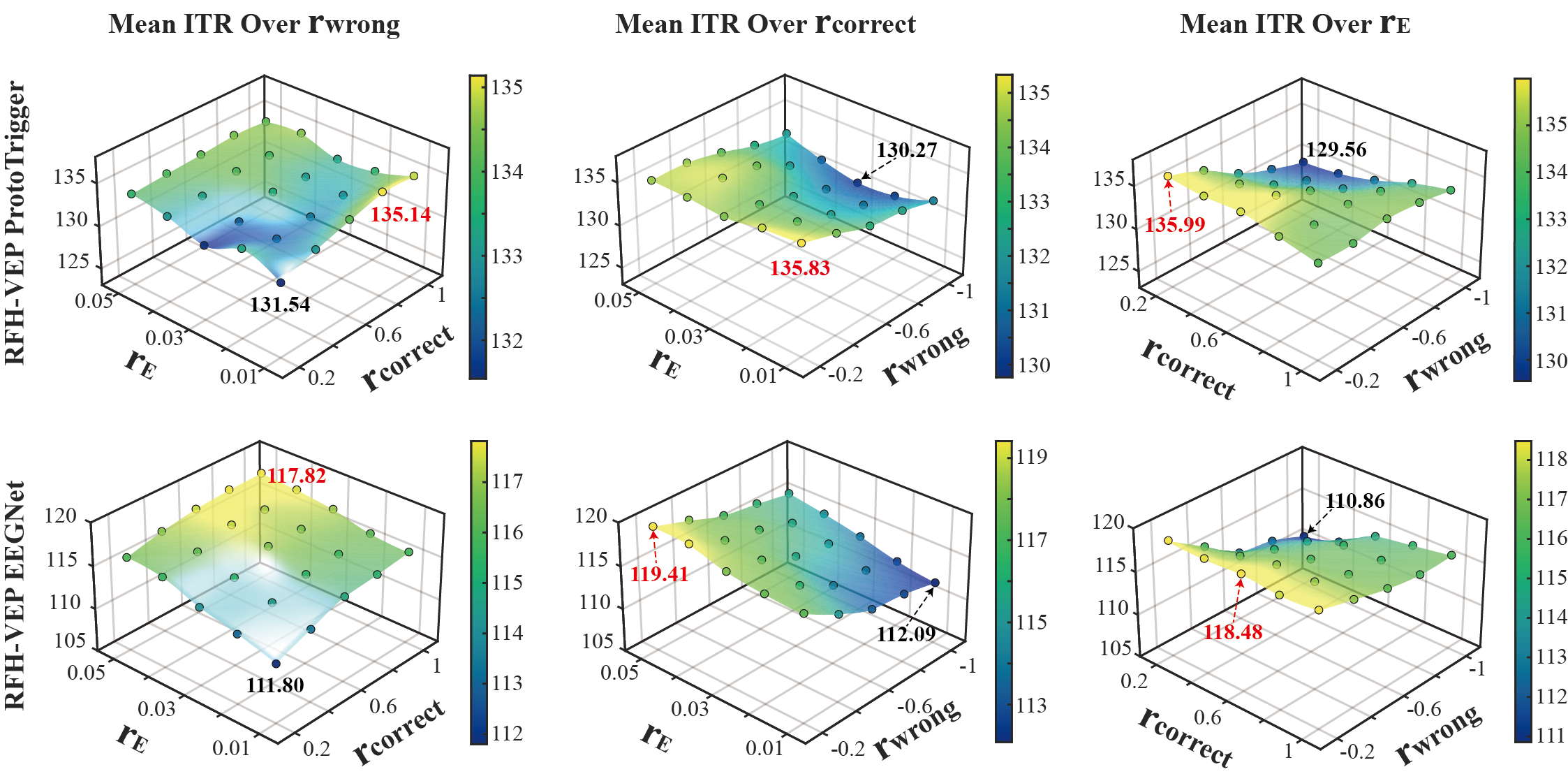}

\caption{RL reward sensitivity analysis on the RFH-VEP dataset. Each column shows the mean ITR surface over two reward parameters after averaging over the remaining parameter, namely $r_{\mathrm{wrong}}$, $r_{\mathrm{correct}}$, and $r_E$ from left to right. The top and bottom rows correspond to ProtoTrigger and EEGNet, the best-performing baseline on RFH-VEP, respectively. Red and black denote the maximum and minimum values, respectively.}
\label{fig:Figure6}
\end{figure*}

The learned state representations were evaluated for their compatibility with RL-based ATDM in terms of cross-length consistency, temporal decision stability, and state discriminability.
Specifically, cross-length consistency was quantified by the displacement of class-wise representation centers across adjacent observation lengths, termed class-center drift (CCD),

\begin{equation}
\mathrm{CCD}
=
\frac{1}{N_{\mathrm{cls}}M}
\sum_{k=1}^{N_{\mathrm{cls}}}
\sum_{t=0}^{M-1}
\left\|
\boldsymbol{\mu}_{k,t+1}
-
\boldsymbol{\mu}_{k,t}
\right\|_{2}
\end{equation}
where $\boldsymbol{\mu}_{k,t}$ denotes the mean state of class $k$ at observation length $T_t$.
Temporal decision stability was quantified using the prediction flip rate (PFR), defined as the frequency of prediction changes between adjacent observation lengths. Here, $N_{\mathrm{te}}$ denotes the number of test trials and $\hat{y}_{i,t}$ denotes the predicted class of the $i$-th test trial at observation length $T_t$,

\begin{equation}
\mathrm{PFR}
=
\frac{1}{N_{\mathrm{te}}M}
\sum_{i=1}^{N_{\mathrm{te}}}
\sum_{t=0}^{M-1}
\mathbb{I}
\left[
\hat{y}_{i,t+1}
\neq
\hat{y}_{i,t}
\right]
\end{equation}
State discriminability was quantified by the average margin between the largest and second-largest class scores, termed the classification margin (CM),
\begin{equation}
\mathrm{CM}
=
\frac{1}{N_{\mathrm{te}}(M+1)}
\sum_{i=1}^{N_{\mathrm{te}}}
\sum_{t=0}^{M}
\left(
z^{(1)}_{i,t}
-
z^{(2)}_{i,t}
\right)
\end{equation}
where $z^{(1)}_{i,t}$ and $z^{(2)}_{i,t}$ denote the highest and second-highest logits, respectively. CCD is motivated by state abstraction and bisimulation theories \cite{kemertas2022approximate}, which suggest that limited representation drift may facilitate stable value estimation and policy learning. PFR reflects temporal decision stability during evidence accumulation, which is important for reliable sequential decision-making \cite{sun2025mooss}. CM characterizes the decision-relevant information preserved in the state, which is required for reward-predictive and predictive state representations \cite{lehnert2020reward}. Therefore, state representations with lower CCD and PFR and higher CM are considered more compatible with RL-based ATDM.

Considering generalizability across EEG paradigms, six data-driven encoders were selected for the state quality evaluation. Following the five-fold cross-validation protocol used in the offline experiments, the metrics were computed from states generated by the pretrained encoder on held-out test data.
Although ProtoTrigger did not achieve the best value for every metric, it maintained competitive cross-length consistency and temporal decision stability while achieving the highest state discriminability across all three datasets. Notably, TCNet achieved lower CCD or PFR on some datasets but consistently exhibited substantially lower CM, indicating weaker state discriminability and decoding performance.

Overall, ProtoTrigger exhibited balanced performance in terms of cross-length consistency, temporal decision stability, and discriminability, suggesting its suitability for state construction in RL-based ATDM.

\subsection{Sensitivity to RL Reward Parameters}

To assess the influence of RL reward configurations on the experimental conclusions, a grid-search analysis was conducted over the extension penalty $r_E$, correct-stopping reward $r_{\mathrm{correct}}$, and wrong-stopping penalty $r_{\mathrm{wrong}}$. Specifically, $r_E$ was varied from $-0.01$ to $-0.05$ with an interval of $-0.01$, $r_{\mathrm{correct}}$ was varied from $0.2$ to $1.0$ with an interval of $0.2$, and $r_{\mathrm{wrong}}$ was varied from $-0.2$ to $-1.0$ with an interval of $-0.2$, resulting in $5\times5\times5$ reward combinations. 

All settings were kept identical to those used in the offline experiments, with only the reward values varied. To visualize the grid-search results, three two-dimensional reward-ITR surfaces were constructed for each method. In each surface, two reward parameters defined the horizontal axes, and the ITR at each grid point was averaged over the five values of the remaining parameter. For each dataset, grid-search results are reported for ProtoTrigger and the second-best method. The complete results are provided in Supplementary Fig.~1.

Reward sensitivity on RFH-VEP is shown in Fig.~\ref{fig:Figure6}. Overall, the ITR of ProtoTrigger remained stable across reward combinations. Across the three pairwise averaged reward surfaces, ProtoTrigger exhibited a narrow mean ITR variation from 129.56 to 135.99~bits/min, indicating low sensitivity to moderate variations in $r_E$, $r_{\mathrm{correct}}$, and $r_{\mathrm{wrong}}$. Moreover, ProtoTrigger consistently outperformed EEGNet, the second-best baseline on RFH-VEP, across the tested reward configurations. EEGNet also exhibited a relatively smooth response surface, with pairwise averaged ITRs ranging from 110.86 to 119.41~bits/min, consistently below those of ProtoTrigger. 

Overall, ProtoTrigger remained superior across the tested reward configurations, demonstrating robustness to reward-parameter variation. Consistent results were observed on the SSVEP and MI datasets (see Supplementary Fig.~1).

\section{Online Validation}

\subsection{Real-time HITL System and Experimental Protocol}

\begin{figure}[h!]
  \centering
  \includegraphics[width=0.95\linewidth]{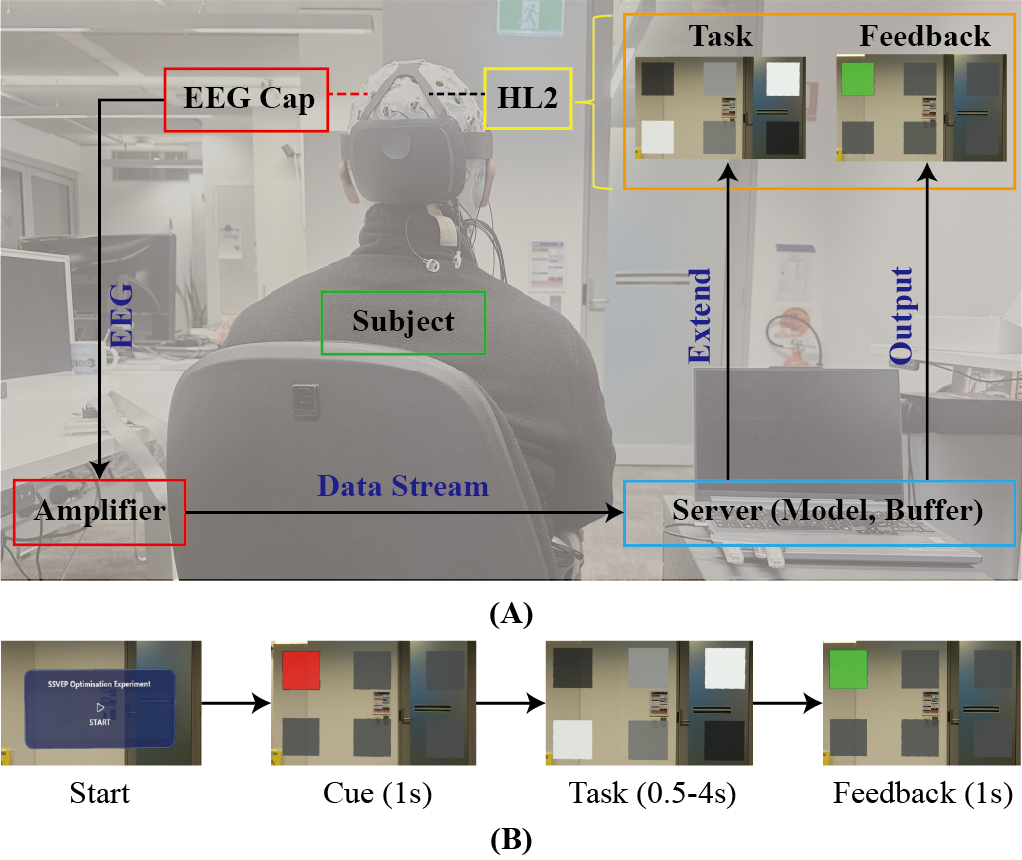}

\caption{Online HITL BCI experimental setup and protocol. (a) Physical setup and data flow of the real-time system. (b) Single-trial procedure.}
\label{fig:Figure7}
\end{figure}

To validate the practical feasibility of ProtoTrigger through an online experiment, a real-time HITL augmented-reality SSVEP (AR-SSVEP) BCI system was developed. As illustrated in Fig.~\ref{fig:Figure7}(a), subjects wore a Microsoft HoloLens~2 (HL2) headset and an EEG cap connected to a LiveAmp~64 amplifier. EEG signals were recorded at 500~Hz and streamed to the processing server through the Lab Streaming Layer (LSL) protocol. The server maintained a real-time EEG buffer and performed preprocessing, ProtoTrigger-based state encoding, RL-based decision-making, and target decoding. At each decision step, the pretrained RL policy selected either an extend action to acquire additional EEG observations or a stop action to output the current prediction. The decoded result was then returned to the AR interface as visual feedback.

The single-trial procedure is shown in Fig.~\ref{fig:Figure7}(b). A six-command AR-SSVEP paradigm was adopted using grayscale luminance-modulated stimuli flickering at frequencies from 7.0 to 9.5~Hz with 0.5-Hz intervals. Each trial was initiated by activation of a virtual Start button through an AR gesture, followed by a 1.0-s cue period during which the target was highlighted in red. Subjects then attended to the flickering target during the task period. Under the ProtoTrigger-ATDM (PTATDM) condition, the initial EEG observation duration was 0.5~s, and the RL policy made an extend-or-stop decision every 0.25~s until a maximum duration of 3.0~s. Two fixed-window (FW) conditions were included using the same pretrained ProtoTrigger encoder and shared prediction head. Predictions were output after fixed observation windows of 1.5~s and 3.0~s under the short- and long-window FW conditions, respectively. Following each prediction, the predicted target was highlighted in green for 1.0~s as visual feedback.

Ten healthy subjects participated in the online experiment. Each subject completed five sessions per condition, each comprising six trials corresponding to the six stimulation targets, for a total of 30 trials per condition. An additional 10 training sessions were conducted with the first subject for model training. The pretrained model was deployed for all subjects without further calibration. All subjects provided written informed consent before the experiment.

\begin{table}[t]
\centering
\caption{Online performance comparison between PTATDM and FWPT strategies. Asterisks indicate significant differences from PTATDM according to paired t-tests (p<0.05).}

\small
\renewcommand{\arraystretch}{1.1}
\setlength{\tabcolsep}{4pt}

\begin{tabular}{lccc}
\toprule
Condition & ACC (\%) & DT (s) & ITR (bits/min) \\
\midrule
\textbf{PTATDM} 
& \textbf{90.33 $\pm$ 7.93} 
& \textbf{1.80 $\pm$ 0.23} 
& \textbf{67.51 $\pm$ 21.95} \\

1.5-s FWPT
& 70.33 $\pm$ 18.42$^{*}$ 
& 1.50 $\pm$ 0.00$^{*}$ 
& 45.45 $\pm$ 27.74$^{*}$ \\

3-s FWPT 
& 91.67 $\pm$ 6.14 
& 3.00 $\pm$ 0.00$^{*}$ 
& 40.47 $\pm$ 7.27$^{*}$ \\
\bottomrule
\end{tabular}

\label{tab:Table3}
\end{table}

\subsection{Online Validation Results}

PTATDM achieved accuracy comparable to 3-s FWPT (90.33\% versus 91.67\%) while reducing DT from 3.00 to 1.80~s, thereby increasing ITR from 40.47 to 67.51~bits/min. Compared with 1.5-s FWPT, PTATDM substantially improved ACC from 70.33\% to 90.33\% with only a modest increase in DT from 1.50 to 1.80~s, resulting in a higher ITR (67.51 versus 45.45~bits/min).
 These results demonstrate that the proposed adaptive strategy reduces decision latency while preserving decoding accuracy, thereby improving the online accuracy--time trade-off and validating the real-time feasibility of ProtoTrigger for ATDM-based BCI.

After completing the experiment, each subject completed the NASA Task Load Index (NASA-TLX) to assess subjective workload under each condition \cite{hart1988development}. As illustrated in Supplementary Fig.~2 and Table~V, PTATDM yielded significantly lower mental, physical, and temporal demands and frustration than 3-s FWPT, while achieving significantly higher perceived performance than 1.5-s FWPT. These results further support the improved user-friendliness of PTATDM.

Computational latency was measured over 10,000 inference runs on an NVIDIA RTX~4070 GPU. RL-based decision-making and final decoding required 0.185~ms and 1.947~ms per trial, respectively, yielding a total inference latency of 2.132~ms, far below the 0.25-s decision interval. The negligible latency supports the practical deployability of the system.

\section{Conclusion and Limitations}

This work proposes ProtoTrigger, a two-stage prototype learning framework designed to enable general EEG state encoding for adaptive temporal decision-making with variable-length neural observations. The prototype-based LFE module is designed to extract effective EEG features from variable-length observations, whereas prototype-based GEA adaptively aggregates and enhances temporally informative evidence for subsequent decision-making. Offline evaluations across three distinct EEG paradigms demonstrated the superiority of ProtoTrigger over conventional encoders, supporting its generalizability across diverse EEG characteristics. The online experiment further validated the feasibility of the proposed encoder in a real-time ATDM BCI system. Overall, ProtoTrigger establishes a unified EEG state encoding framework for more efficient and user-friendly real-time BCI interaction.

Despite these promising results, several limitations remain. First, online validation was limited to a relatively small-scale AR-SSVEP experiment, and the real-time applicability of ProtoTrigger to other BCI systems requires further investigation. Second, the methodological development primarily focused on the EEG state encoder, while the downstream policy adopted a standard DQN and was trained separately. Joint optimization of state encoding and policy learning remains to be explored.

\bibliographystyle{IEEEtran}
\bibliography{reference}
\end{document}